\documentstyle[twocolumn,aps]{revtex}
\def\go{\mathrel{\raise.3ex\hbox{$>$}\mkern-14mu
             \lower0.6ex\hbox{$\sim$}}}
\def\lo{\mathrel{\raise.3ex\hbox{$<$}\mkern-14mu
             \lower0.6ex\hbox{$\sim$}}}
\def\be{\begin{equation}}
\def\ee{\end{equation}}

\begin{document}
\tightenlines
\draft
\title{
Innermost Stable Circular Orbit of Inspiraling \\
Neutron-Star Binaries:  Tidal Effects, Post-Newtonian Effects \\
and the Neutron-Star Equation of State }
\author{Dong Lai  and  Alan G. Wiseman}
\address{Theoretical Astrophysics, 130-33,
California Institute of Technology\\
Pasadena, CA 91125\\
{\rm E-mail: dong@tapir.caltech.edu;  agw@tapir.caltech.edu}}
\date{\today} 
\maketitle 
\begin{abstract}

We study how the neutron-star equation of state affects
the onset of the dynamical instability in the equations
of motion for inspiraling neutron-star binaries near coalescence.
A combination of relativistic effects
and Newtonian tidal effects cause the stars to begin
their final, rapid, and dynamically-unstable plunge to merger
when the stars are still well separated and the orbital frequency is
$\approx$ 500 cycles/sec ({\it i.e.} the gravitational wave frequency
is approximately 1000 Hz).
The orbital frequency at which the dynamical instability occurs
({\it i.e.} the orbital frequency at the innermost stable circular orbit)
shows modest sensitivity to the neutron-star equation of state
(particularly the mass-radius ratio, $M/R_o$, of the stars).
This suggests that information about the equation of state of nuclear
matter is encoded in the gravitational waves emitted just 
prior to the merger.

\end{abstract}
\bigskip
\pacs{PACS Numbers: 97.80.Fk, 04.25.Dm, 04.40.Dg, 97.60.Jd}

\section{\bf INTRODUCTION}

Binary neutron star systems which are spiraling toward their final
coalescence under the dissipative influence of gravitational 
radiation reaction forces are the primary targets for detection of
gravitational waves by interferometric gravitational wave
detectors such as LIGO and VIRGO\cite{Abramovici92,Thorne95}.
Extracting
the gravitational waves from the detector noise and 
making use of the information encoded in the signals will require a
thorough knowledge of the expected waveforms produced by these
binaries\cite{Thorne95,Cutler93}.
In this paper we explore the effect of the neutron-star equation of
state on the orbital evolution and gravitational wave
emission of binaries just prior to merging.
Specifically, we show that a combination of post-Newtonian
(relativistic) effects and Newtonian tidal effects
(which depend on the equation of state) conspire to induce 
a dynamical instability in the orbital motion, which
causes the plunge to final coalescence to begin somewhat
sooner -- and to proceed somewhat faster -- than it would simply under
the influence of the dissipative radiation reaction force.
Thus the motion of the bodies during the late stages of binary
inspiral depends on the structure of the neutron stars. 
Consequently, the gravitational waveform emitted during this short
portion of the final coalescence will be imprinted
with information about the nuclear equation of state.

During the final $\sim 10$ minutes or the last $\sim 8000$ orbits of a
neutron star binary inspiral, the orbital frequency increases from
about 5 Hz on up to a cutoff of a few hundreds to a thousand
Hertz (roughly corresponding to the orbital frequency when the 
final plunge begins). Thus the gravitational wave frequency (twice the
orbital frequency for the dominant quadrupole radiation) {\it chirps}
through the LIGO detector bandwidth during
this period \cite{Abramovici92}.
The evolution of the binary in these last few minutes
of the inspiral is very sensitive to a number of relativistic 
effects, such as gravitational-wave tails and
spin-orbit coupling (dragging of inertial frames). 
The gravitational waveform emitted by the binary in this
portion of the coalescence, the adiabatic inspiral, is currently 
being extensively studied \cite{bdiww,biww,ww}.
During most of this inspiral phase the neutron stars can be 
treated as simple point masses because the effects associated with 
the finite stellar size turn out to be small: 
(i) The neutron star has too small a viscosity to allow for angular
momentum transfer from the orbit to the stellar spin
via viscous tidal torque\cite{Bildsten92,Kochanek92};
(ii) The effect of the spin-induced quadrupole is negligible
unless the neutron star has rotation rate close to the break-up
limit\cite{Bildsten92,LRS94}; (iii) Resonant excitations of neutron
star internal modes (which occur at orbital frequencies less than
$100$ Hz) produce only a small change in the orbital phase due to the weak
coupling between the modes and the tidal 
potential\cite{Lai94,Reisenegger94,Shibata94};
(iv) The correction to the equation of motion from the (static) tidal
interaction is of order $(R_o/r)^5$ (where $R_o$ is the neutron star
radius, $r$ is the orbital separation\cite{notation}), 
which is negligible except when $r$ is smaller than a few stellar
radii. Since $R_o\simeq 5M$  
for a typical neutron star of mass $1.4M_\odot$ and radius $10$ km,  
the tidal effect is essentially a (post)$^5$-Newtonian 
correction \cite{quad}.
The expression for the phase error induced by the tidal effect is given in
Ref.\cite{LRS94}.
The fact that the evolution of the binary system as it sweeps through
the low frequency band of the detector is insensitive to 
finite-size effects means that the measurement of the
inspiral waveform will allow us to probe cleanly into
the intricate structure of general relativity, and to test
whether general relativity is the correct theory of 
gravity \cite{lucsathya,cliffscalar}. Moreover, some of
the parameters of the binary system, such as the masses of the stars, can
be determined with reasonable accuracy\cite{finnchernoff,cutlerflanagan}.
However, the waveform's lack of dependence on the finite size of the
objects during the most of the adiabatic inspiral 
also implies that information about the internal structure of the
neutron star is only imprinted on the radiation emitted
just prior to coalescence when the orbital radius is small.

Indeed at small orbital separations,
tidal effects are expected to be very important.
In a purely {\it Newtonian} analysis, the interaction potential
between star $M'$ and the tide-induced quadrupole of $M$, 
$V_{tide}\sim -M'^2R_o^5/r^6$, increases with decreasing $r$. 
The potential becomes so steep that a dynamical instability 
develops, accelerating the coalescence at small orbital 
radius. This Newtonian instability has been fully explored 
using semi-analytic models in Ref.\cite{LRS94} and
Ref.\cite{LS95} (hereafter referred to as LS).
It has also been examined numerically in Refs.\cite{Rasio,Centrella}.

However, a purely Newtonian treatment of the binary at small separation
is clearly not adequate, as general relativistic effects
will also be important in this regime; and general relativistic
effects can also make the orbit unstable.
For example, a test particle in circular orbit around a
Schwarzschild black hole will experience an ``innermost stable circular
orbit'' at $r_{\rm isco}=6M$ (or $5M$ in harmonic coordinates). This
unstable behavior is caused by higher-order relativistic corrections
included in the Schwarzschild geodesic equations of motion.
For computing the orbital evolution of two neutron stars of
comparable mass near coalescence, the test-mass limit is obviously
inadequate. 
In order to explore the orbital instability for 
such systems, Kidder, Will and Wiseman 
\cite{KWW93} (hereafter referred as KWW) 
developed {\it hybrid} equations of motion.
These equations augment the the Schwarzschild geodesic equations
of motion with the finite-mass terms of the (post)$^{5/2}$-Newtonian
equations of motion.
Including these finite mass terms in the equation of motion
moved the innermost-stable-circular-orbit radius farther out (in units
of the total mass).\footnote{See Wex and Sch\"afer\cite{Wex93}
for a critique and an alternative
construction. Their post-Newtonian calculation suggests that
the innermost stable orbit may occur at an even greater separation.}
In this paper, we augment the hybrid equations
with contributions due to the tidal deformation of the stars. 
In a nutshell, the work presented here combines the Newtonian
tidal analysis of LS \cite{LS95} with the relativistic point-mass
analysis of KWW \cite{KWW93} to yield a more complete picture of the
neutron-star coalescence prior to merging. 

We note, that unlike a test-particle around a Schwarzschild
black hole, the very notion of ``innermost stable circular orbit'' 
is poorly defined for objects of comparable mass.
After all, in the relativistic regime the binary orbit will be decaying 
rapidly due to radiation reaction; thus the orbit 
is not circular, but rather a decaying spiral. 
In order to give a semi-quantitative definition of ``innermost
stable circular orbit'' we use the artifice of ``shutting off''
all the dissipative terms in the equation of motion and
looking for the point where the solutions of the
remaining non-dissipative equations become dynamically unstable.
The use of hybrid equations of motion
augmented with the tidal terms
allows us to map out the dependence of 
the critical radius $r_{\rm isco}$, or the corresponding orbital 
frequency $f_{\rm isco}$, for a wide range of allowed neutron star equations
of state (parametrized by radius and effective polytropic index; see
Figure 1). We believe that clearing up such dependence is
important, 
and this analysis provides a benchmark with which comparisons can be 
made with future numerical results. 
{\it Indeed, an important point we wish
to make in this paper is that neither 
relativistic (post Newtonian) effects nor Newtonian tidal effects
can be neglected near the instability limit, and the critical
frequency can be much lower than 
the value obtained when only one of these effects are included.}

\begin{figure}[t]
\special{hscale=42 vscale=42 hoffset=0.0 voffset=-285.0
         angle=0.0 psfile=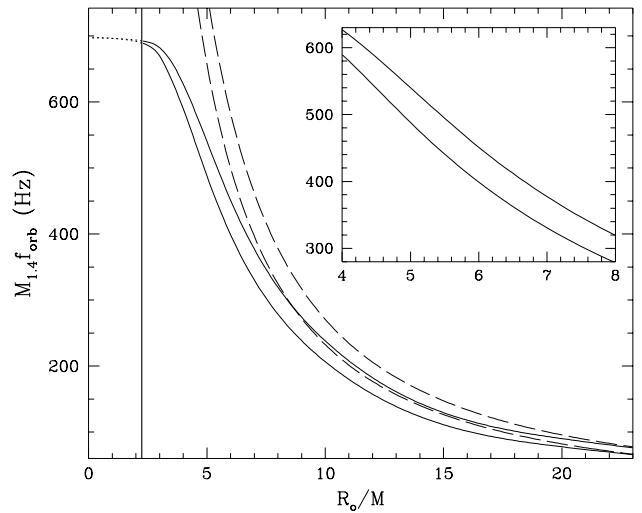}
\vspace*{2.8in}
\caption[Fig. 1]{
The critical orbital frequency (at the inner-most stable orbit)
as a function of the ratio of the neutron star radius $R_o$ and
mass $M$. The solid curves show the results including both 
relativistic and
tidal effects (the lower curve is for $\Gamma=3$ while the upper one 
is for $\Gamma=2$), the dashed curves are the Newtonian limit
given by Eqs.~(8)-(9). The vertical line corresponds to
$R_o/M=9/4$, the minimum value for any physical neutron star. 
The insert is a close-up for the nominal range of $R_o/M=4-8$
as given by all the available nuclear EOS's. 
Two curves within the insert should bracket all the physical 
values of $f_{\rm isco}$.}
\end{figure}

The main results of our analysis are summarized in Figure 2 and 3.
Figure 2 shows that the rate of radial infall for stars
near coalescence is substantially 
underestimated if one models the coalescence as
a Newtonian circular-orbit decaying solely under the
influence of radiation reaction (top dotted curve).
In other words, the rate of coordinate infall is substantially
enhanced by the non-dissipative terms.
Somewhat more relevant for observational purposes, Figure 3
shows that the number of orbits (or gravitational wave cycles) 
per logarithmic frequency interval is substantially reduced
by the unstable collapse of the orbit.
Both plots show modest sensitivity to the equation of state.

\begin{figure}[t]
\special{hscale=52 vscale=55 hoffset=-42.0 voffset=-365.0
         angle=0.0 psfile=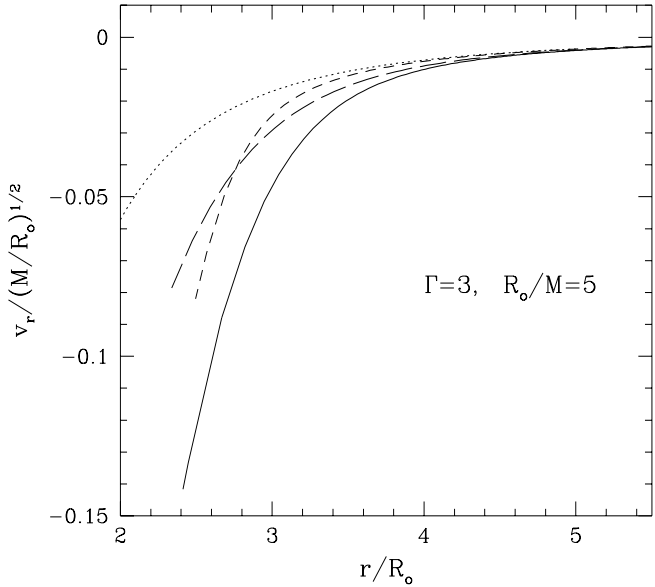}
\vspace*{3.1in}
\caption[Fig. 2]{
The radial infall coordinate velocity during binary coalescence, with 
$M=M'=1.4M_\odot$, $R_o/M=5$, $\Gamma=3$, all calculated using 
$2.5PN$ radiation reaction.    
The solid line is the result including relativistic and tidal effects,
the short-dashed line includes only tidal effects, the long-dashed
line includes only relativistic effects.
The dotted line is the point mass ``Newtonian'' result. 
}
\end{figure}

\begin{figure}[t]
\special{hscale=52 vscale=55 hoffset=-50.0 voffset=-360.0
         angle=0.0 psfile=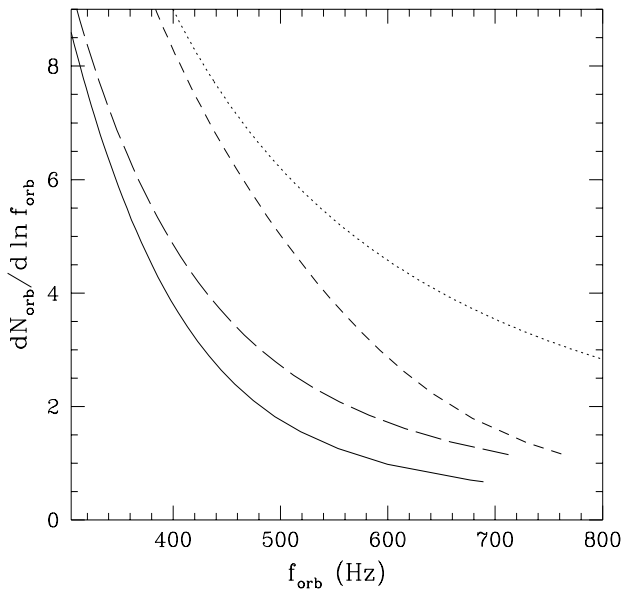}
\vspace*{3.1in}
\caption[Fig. 3]{
The number of orbits the binary spends per logarithmic
frequency. The labels are the same as in Fig.~2. 
}
\end{figure}

The remainder of the paper is organized as follows:
In section II we present our equations of motion. 
In section III we examine the orbital stability using the non-dissipative 
portion of the equations of motion,
and thus identify the location of the ``innermost stable circular orbit''.
In section IV we include the dissipative terms that were omitted in
the analysis of section III, and evolve the full equations of motion.
In section V we briefly discuss the relevance of our results to 
numerical hydrodynamic calculations and to gravitational
wave signal analysis.

\section{\bf EQUATIONS OF MOTION INCLUDING 
TIDAL AND GENERAL RELATIVISTIC EFFECTS}

Consider a binary containing two neutron stars of mass $M$ and $M'$,
each obeying a polytropic equation of state
$P=K\rho^\Gamma$. We use the compressible ellipsoid model 
for binary stars developed in LS\cite{LS95}. Basically, 
we model the tidally deformed neutron star
as an ellipsoid, with internal density profile 
similar to that of a spherical polytrope. 
The dynamics of such a neutron star (so called 
Riemann-S ellipsoid) is characterized by the three
principal axes ($a_1,a_2,a_3$ for star $M$ and $a_1',a_2',a_3'$
for star $M'$), the angular velocity ($\Omega$ and $\Omega'$)
of the ellipsoidal figure about a principal axis (perpendicular to 
the orbital plane) and the internal motion of the 
fluid with uniform vorticity. The non-zero internal fluid motion is
necessary because the binary neutron stars are not expected
to corotate with the orbit due to rapid orbital decay and small
viscosity\cite{Bildsten92,Kochanek92}. Although the Newtonian tidal
interaction between the neutron stars can be treated
exactly in the linear regime using mode decomposition\cite{Lai94},
the ellipsoid model has the advantage that it can be extended to 
the nonlinear regime at small orbital radii, when the tidal
deformation of the star becomes significant. 

The Newtonian dynamical equations for the binary neutron stars
as derived in LS include the familiar
Newtonian ($1/r^2$) force-law for point-masses orbiting one another;
they further contain Newtonian terms involving finite 
size (tidal) effects. A post-Newtonian
treatment of the tidal problem would give the relativistic
corrections to these terms, namely
the standard point-mass, post-Newtonian corrections to the
equations of motion, as well as relativistic corrections
to the quadrupole moment and corrections 
due to higher moments the bodies. (See Appendix F of \cite{WW96}.)
To insure that our equations of motion at least agree with 
the known post-Newtonian, point-mass equations we augment these
Newtonian equations of motion with the hybrid equations of
KWW. However, we use only the Newtonian equations to describe the
evolution of the neutron stars structure ($a_i$ and $a_i'$) and the
fluid motion (the figure rotation rate and the internal vorticity)
within the stars. These are given by Eqs.~(2.18)-(2.22) of LS. 
In other words, we neglect the relativistic corrections to the fluid
motion, self-gravity and tidal interaction. 
These corrections are secondary effects and should not modify the 
{\it orbital dynamics} appreciably (e.g., the Newtonian tidal
interaction between the two stars scales approximately as
$M^2R_o^5/r^6$, and its relativistic correction is of order $M/r$
smaller). As noted before, the tidal interaction
enters the Newtonian potential as a correction of 
$O[(R_o/r)^5]\sim O[(M_t/r)^5]$, in effect, as a (post)$^5$-Newtonian
term. Therefore, by not including relativistic corrections to 
these tidal terms, we are merely omiting terms
which are of (post)$^6$-Newtonian order.
In fact, the largest error comes
from neglecting the post-Newtonian correction (of order $M/R_o$) to
the internal stellar structure (see Sec.~III.B for an estimate of its
effect on $f_{\rm isco}$). The relativistic corrections to the
orbital motion, however, are very important.
Our equations of orbital motion can be assembled from
Eqs.~(2.23)-(2.24) of LS and from Eqs.~(1.2)-(1.3) of KWW:
\begin{eqnarray}
\ddot r &=& r{\dot\theta}^2-{M_t\over r^2}(A_H+B_H\dot r)\nonumber\\
&&-{3\kappa_n\over 10}{M_t\over r^4}\left[a_1^2(3\cos^2\alpha-1)
+a_2^2(3\sin^2\alpha-1)-a_3^2\right] \nonumber\\
&&-{3\kappa_n'\over 10}{M_t\over r^4}\left[a_1'^2(3\cos^2\alpha'-1)
+a_2'^2(3\sin^2\alpha'-1)-a_3'^2\right] \nonumber\\
&&-{M_t\over r^2}(A_{5/2}+B_{5/2}\dot r)
-{32\over 5}r\bigl[\Omega^5(I_{11}-I_{22})\sin 2\alpha \nonumber\\
&&+\Omega'^5(I_{11}'-I_{22}')\sin 2\alpha'\bigr],
\\
\ddot\theta &=& -{2\dot r\dot\theta\over r}
-{M_t\over r^2}B_H\dot\theta\nonumber\\
&&-{3\kappa_n\over 10}{M_t\over r^5}(a_1^2-a_2^2)\sin 2\alpha
-{3\kappa_n'\over 10}{M_t\over r^5}(a_1'^2-a_2'^2)\sin 2\alpha' \nonumber\\
&&-{M_t\over r^2}B_{5/2}\dot\theta
-{32\over 5}\bigl[\Omega^5(I_{11}-I_{22})\cos 2\alpha \nonumber\\
&&+\Omega'^5(I_{11}'-I_{22}')\cos 2\alpha'\bigr],
\end{eqnarray}
where $M_t=M+M'$ is the total mass, $\alpha$ ($\alpha'$) is
the misalignment angle between the tidal bulge of $M$ ($M'$)
and the line joing the two masses, $\kappa_n,\kappa_n'$ are 
dimensionless structure constants depending on the mass concentration
within the stars.
In Eqs.~(1) and (2) the last two lines contain the ``dissipative''
terms due to gravitational radiation reaction.
The quantities $A_H$, $B_H$, $A_{5/2}$, $B_{5/2}$, 
which include the ``hybrid'' corrections to the equation of motion,
are given by 
\begin{eqnarray}
A_H &=& {1-M_t/r\over (1+M_t/r)^3}-\left[{2-M_t/r\over
1-(M_t/r)^2}\right]{M_t\over r}\dot r^2+v^2\nonumber\\
&&-\eta\left(2{M_t\over r}-3v^2+{3\over 2}\dot r^2\right)
+\eta\biggl[{87\over 4}\left({M_t\over r}\right)^2\nonumber\\
&&+(3-4\eta)v^4+{15\over 8}(1-3\eta)\dot r^4
-{3\over 2}(3-4\eta)v^2\dot r^2\nonumber\\
&&-{1\over 2}(13-4\eta){M_t\over r}v^2
-(25+2\eta){M_t\over r}\dot r^2\biggr] 
\\
B_H &=& -\left[{4-2M_t/r\over 1-(M_t/r)^2}\right]\dot r
+2\eta\dot r-{1\over 2}\eta\dot r\biggl[(15+4\eta)v^2\nonumber\\
&&-(41+8\eta){M_t\over r}-3(3+2\eta)\dot r^2\biggr], 
\\
A_{5/2} &=& -{8\over 5}\eta{M_t\over r}\dot r\left(18v^2
+{2\over 3}{M_t\over r}-25\dot r^2\right),
\\
B_{5/2} &=& {8\over 5}\eta{M_t\over r}\left(6v^2
-2{M_t\over r}-15\dot r^2\right) \; ,
\end{eqnarray}
where $v^2=\dot r^2+r^2\dot\theta^2$, 
$\eta=\mu/M_t$ and $\mu=MM'/M_t$. 
Also the multipoles moments can be expressed as
\be
I_{ii}=\kappa_nMa_i^2/5,~~~~~
I_{ii}'=\kappa_n'M'a_i'^2/5 \; .
\ee
Note that in Eqs.~(1)-(2), we have also included the leading order
radiation reaction forces due to tidal deformation. 

Admittedly, this is not a consistent post-Newtonian expansion of the 
true equations of motion; however it is correct in several important
limiting cases: (i) In the limit that 
$a_i \rightarrow 0$ and $a_i' \rightarrow 0$ and the 
limit $\eta \rightarrow 0$, we recover 
the {\it exact} Schwarzschild equation of motion.
(ii) In the point-mass limit ($a_i\rightarrow 0$ and $a_i'\rightarrow 0$)
we recover the hybrid equations given in KWW.
KWW presented an argument that suggested
that the higher-order, $\eta$-dependent, (post)$^3$-Newtonian
-- as yet uncalculated -- corrections to these equations have only a modest
effect on the equations of motion. See Figure 6 of Ref.\cite{KWW93}. 
However, until these terms are calculated it is unclear just how
large an effect they will have on the location of the innermost
stable orbit.
(iii) In the non-relativistic limit we recover the equations of motion
given in LS. These equations contain the dominant contributions to the
equations of motion due to the finite sizes of the objects.

Note that although Eqs.~(1) and (2) make reference to the orbital
radius $r$, we are always aware that this is a gauge dependent quantity
and of little meaning for a distant observer.
Observationally, the more meaningful quantity is the orbital frequency
as measured by distant observers, and we shall use frequency 
rather the radius in presenting most of our results.

\section{\bf INSTABILITY OF THE NON-DISIPATIVE EQUATIONS OF MOTION}

\subsection{Method to Determine the Stability Limit}

We now form a set of non-dissipative equations of motion,
by simply discarding the gravitational radiation reaction terms 
given in the last two lines of Eq.~(1) and Eq.~(2).
These non-dissipative dynamical equations admit equilibrium solutions, 
which are obtained by setting $\dot r=\ddot
r=\ddot\theta=\dot\Omega_{orb}=\alpha=\alpha'=0$ as well as
$\dot a_i=\dot a_i'=0$. For a given $r$, the evolution equations for
the neutron star structure reduce to a set of algebraic equations 
for $a_i$ and $a_i'$, while the orbital equation (2) gives the orbital
frequency $\Omega_{orb}$. These equations are solved using a
Newton-Raphson method, yielding an equilibrium binary model. Thus a
sequence of binary models parametrized by $r$ can be constructed. 

To determine the stability of the orbit of a binary model, 
we simply use the equilibrium parameters as initial conditions 
for our non-dissipative equations of motion. We add a small perturbation
to the equilibrium model and let the system evolve. 
In this way we locate the critical point of the dynamical equations,
corresponding to the dynamical stability limit of the equilibrium
binary or the inner-most stable circular orbit: 
for $r>r_{\rm isco}$, the binary is stable, and the
system oscillates with small amplitude about the initial
configuration; for $r<r_{\rm isco}$, the binary is unstable, and 
the perturbation grows, leading to the swift merger of the
neutron stars even in the absence of dissipation.

\subsection{Results}

For concreteness, we present results only for binary neutron stars 
with equal masses ($M=M'$), both having zero spin at 
large orbital separation, although our equations are adequate
to treat the most general cases\cite{LS95}.

The polytropic relation $P=K\rho^\Gamma$ provides a useful
parametrization to the most general realistic nuclear equation of
state (EOS). Since the radius $R_o$ of the nonrotating neutron star
of mass $M$ is uniquely determined by $K$ and $\Gamma$, we can 
alternatively use $R_o/M$ and $\Gamma$ to characterize the EOS.
For a canonical neutron star with mass $M=1.4M_\odot$, all EOS
tabulated in\cite{Arnett77} give $R_o/M$ in the range of $4-8$,
while modern microscopic nuclear calculations typically give
$R_o/M=5$\cite{Wiringa88}. For a given $R_o/M$, the polytropic index
$\Gamma$ specifies the mass concentration within the star. Except for
extreme neutron star masses ($M\lo 0.5M_\odot$ or $M\go 1.8M_\odot$)
typical values of $\Gamma$ lie in the range of $\Gamma=2-3$
\cite{LRS94}.

In Table I, we list the physical properties of 
the equilibrium binary neutron stars at the dynamical stability 
limit for several values of $R_o/M$ and $\Gamma=3$. 
In Figure 1, the orbital frequency $f_{\rm isco}$ 
is shown as a function of $R_o/M$ for $\Gamma=2$ and
$\Gamma=3$. Clearly, in the limit of $R_o/M\rightarrow 0$, 
$f_{\rm isco}$ approachs the point mass result $f_{\rm
isco}=697M_{1.4}^{-1}$
Hz obtained in KWW\footnote{KWW\cite{KWW93}
give a correct expression for $r_{\rm isco}/M_t$, but incorrectly give
$f_{\rm isco}$=710 Hz due to a numerical error.}. 
In the non-relativistic limit
we recover the pure Newtonian result\cite{LRS94,LS95}:
\begin{eqnarray}
f_{\rm isco} &=& 657M_{1.4}^{-1}(5M/R_o)^{3/2}~({\rm Hz})
~~~~~(\Gamma=3),\\
f_{\rm isco} &=& 766M_{1.4}^{-1}(5M/R_o)^{3/2}~({\rm Hz})
~~~~~(\Gamma=2).
\end{eqnarray}
{\it For typical neutron star radius $R_o/M=5$, the critical 
frequency ranges from $488$ Hz (for $\Gamma=3$) 
to $540$ Hz (for $\Gamma=2$), while both the pure Newtonian (with tides)
calculation and the pure point-mass hybrid calculation give a result
$30-40\%$ larger}. There are two physical causes for the reduction
in $f_{\rm isco}$: (i) The binary becomes unstable at larger 
orbital separation due to the steepening of the interaction potential 
from both tidal and relativistic effects;
(ii) For a given orbital radius (itself a gauge dependent quantity),
the post-Newtonian orbital frequency as measured by an observer at
infinity is smaller than the Newtonian orbital
frequency\footnote{In the case of equal masses,
at first post-Newtonian order
$\Omega_{orb} = \Omega_{Kepler}[1 - (11/8)(M_t/r)]$.
See Ref.\cite{bdiww}.}.
We conclude that to neglect either the 
tidal effects or the relativistic effects can lead to large error in
the estimated critical frequency.

Except for the intrinsic uncertainties associated with 
the hybrid equations of motion\cite{KWW93,Wex93}, the main 
uncertainty in our determination of $f_{\rm isco}$ comes from 
neglecting post-Newtonian corrections to (i) the stellar structure and
(ii) the the tidal potential. 
The first correction {\it decreases} the tide-induced quadrupole; 
the fractional change is of order $-M/R_o$. 
The second {\it increases} the quadrupole by a fraction of order
$M'/r$. We can estimate how much $f_{\rm isco}$ is modified by these
two corrections. 
In Newtonian theory, $r_{\rm isco}$ is approximately determined by 
the condition $MM'/r\sim M'^2R_o^5/r^6$. Including the relativistic
corrections
this condition becomes $MM'/r\sim M'^2R_o^5/r^6(1-\delta)$, where
$\delta\sim [O(M/R_o)-O(M'/r)]\lo 20\%$. Thus the change
in $f_{\rm isco}$ due to these two effects is 
$\Delta f_{\rm isco}/f_{\rm isco}\simeq
0.3\,\delta\lo 6\%$, i.e., the critical frequency increases by 
a few percent\cite{Wilsonnote}. 

As emphasized in Sec.~I, the critical radius (or critical frequency),
at which the non-dissipative equations become dynamically unstable,
is meaningful only in the sense that when $r<r_{\rm isco}$, the binary 
will coalesce on dynamical (orbital) timescale even
in the absence of dissipation. In the realistic situation, 
the dissipative radiation reaction forces will also be rapidly driving
the binary to coalescence. Therefore to determine the significance of the
dynamical instability we must compute the orbital evolution with
the full equations of motion -- including the radiation reaction.

\section{\bf ORBITAL EVOLUTION PRIOR TO MERGER}

We now include the dissipative radiation reaction forces in our analysis.
In this case the plunge will be driven by both the
dissipative, as well as the non-dissipative effects associated with 
the steepening potential (both the tidal potential and the
relativistic potential). 
But what effect is dominant? 
In order to numerically investigate this question,
we choose a specific system with $M=M'=1.4M_\odot$, $R_o/M=5$ and
$\Gamma=3$. The orbital evolution begins when the stars are well
outside the innermost stable circular orbit limit. 
We consider four different inspiral scenarios. 
(i) A purely dissipative inspiral: a system of point masses subject only
to a Newtonian ($1/r^2$) force and (post)$^{5/2}$-Newtonian
radiation reaction force. In this case the infall rate is given by
$v_r=\dot r=-(64/5)\eta (M_t/r)^3$. 
[Specifically, we set
$A_H = 1$  and $B_H = a_i = a_i' = I_{kk} = I_{kk}' = 0$ in
Eqs.~(1)-(2).]
This is depicted by the dotted
curve in Figure 2 and 3. 
(ii) A purely relativistic plunge in which 
we neglect the tidal effects 
[Specifically we set $a_i=a_i'= I_{kk} = I_{kk}' =0$ in Eqs.~(1)-(2)].
This relativistic case is depicted by the long-dashed curve in Figure
2 and 3. 
(iii) A tidally enhanced plunge: we include only the
Newtonian terms in Eqs.~(1)-(2) and the radiation reaction force. 
[Specifically we set $A_H = 1$ and $B_H = 0$ in Eqs.~(1)-(2).]
This case is depicted by the short-dashed curve in Figure 2 and 3. 
(iv) Finally, we evolve the complete dynamical equations including all
terms in Eqs.~(1)-(2); this is depicted by the solid curve in Figure 2
and 3. 
Each intergration is terminated when the surfaces of the stars
touch, i.e., at $r\simeq 2a_1$ (for the point-mass problem, 
the calculation is terminated at $r\simeq 2R_o$).

In Figure 2 we clearly see that the non-dissipative effects
-- tidal and relativistic -- substantially increase the rate of 
infall. The radial velocity at binary contact is comparable to the
tangential velocity. We also note that the radial coordinate 
velocity is a gauge dependent quantity;
therefore our only intent in using it in Figure 2 is to
convey the general trend that the rate of infall is enhanced by the
dynamical instability.

Figure 3 shows the number of orbits the binary spends per logarithmic 
frequency. In the simplest point-mass,
Newtonian-plus-radiation-reaction case
[case (i) above], the result can be calculated
analytically
\begin{eqnarray}
dN_{orb}/d\ln f_{orb} &=&(5/192\pi)\mu^{-1}M_t^{-2/3}
(2\pi f_{orb})^{-5/3}\nonumber\\
&=&1.95\times 10^5(M_{1.4}f_{orb}/{\rm Hz})^{-5/3},
\end{eqnarray}
which gives $6$ cycles at $f_{gw}\simeq
2f_{orb}=1000$ Hz. In contrast, the tidal and relativistic effects
reduce this number to less than $2$.

Figure 4 shows the wave energy emitted around a given frequency,
$dE_{gw}/d\ln f_{orb}=(\Omega_{orb}/\dot\Omega_{orb})\dot E_{gw}$,
where $\dot E_{gw}$ is calculated using the simple quadrupole
radiation formula. The Newtonian plus radiation reaction result is
$dE_{gw}/d\ln f_{orb}=1.63\times 10^{-3}(f_{orb}/
{\rm Hz})^{2/3}M^2/R_o$. We see that the radiation power near 
contact becomes much smaller. 
Note that $dE_{gw}/d\ln f_{orb}$ calcuated in this way is not 
exactly the energy power spectrum, which must be obtained
from the Fourier transform of the waveform\cite{Kennefick96}; 
however, it does provide a semi-quantitative feature
of the full analysis; in particular, the dip in the 
$dE_{gw}/d\ln f_{orb}$ curve around $600$ Hz results
from the dynamical instability of the orbit (see also 
Refs.\cite{Centrella,Ruffert}, although the calculations presented 
there are purely Newtonian). 

\begin{figure}[t]
\special{hscale=52 vscale=55 hoffset=-50.0 voffset=-375.0
         angle=0.0 psfile=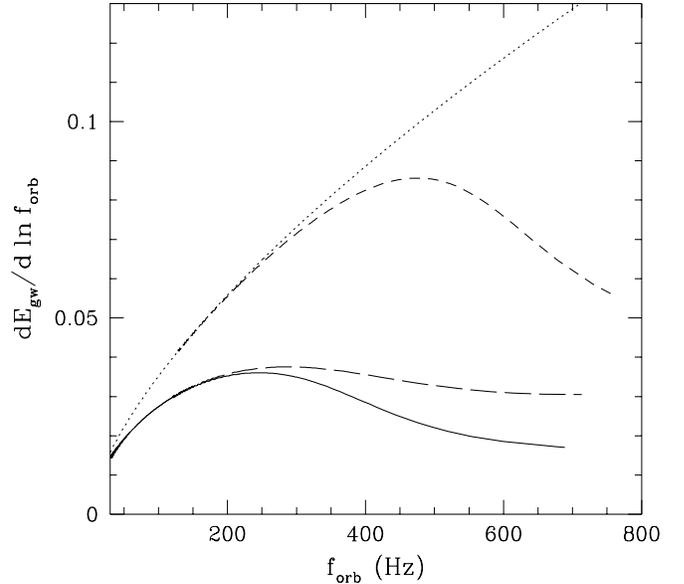}
\vspace*{3.1in}
\caption[Fig. 4]{
The quadrupolar gravitational energy emitted near a given frequency.
The labels are the same as in Fig.~2.
}
\end{figure}

\section{\bf DISCUSSION}

A number of authors have tried to define and locate the
innermost stable circular orbit for relativistic coalescing
systems of comparable masses
\cite{Wex93,Wilson95,clarkeardley,blackburndetweiler,cook}
in order to characterize the final moments of a binary coalescence
(See \cite{eardleyhirschnamm} for a discussion).
The results of the various analyses are not converging to an agreed
answer. Obviously, the precise nature of the final coalescence of two
neutron stars will only be determined by a numerical simulation
using full general-relativistic hydrodynamics.
However, the present analysis does point to two intersting features
to look for in a full numerical treatment:
(i) To get even a qualitative picture of the
coalescence, it is necessary to begin the numerical evolution
when the stars are still well separated, {\it i.e.}  before
the onset of the orbital dynamical instability.
The instability -- the plunge -- causes the coalescence
to proceed much more swiftly than a coalescence driven
solely by radiation reaction; thus the actual coalescence may differ
qualitatively from one computed with a simple 
radiation-reaction driven inspiral.
The final coalescence may be more of a splat,
than the slow winding together of the stars\cite{Nakamura}.
(ii) The instability results from both the tidal effects and 
the relativistic corrections in the equations
of motion.  KWW showed that there
is no instability in the {\it first} post-Newtonian relativistic
equations of motion; the instability does not show up until at least
second post-Newtonian order. 
Therefore, in order for numerical simulations to see the
effects of the relativistic unstable plunge,
it will probably require the use of at least second-order, post-Newtonian
hydrodynamic code.  KWW also showed that the location of
the dynamic instability does not converge very rapidly as
one increases the post-Newtonian order of the approximation.
[This fact led KWW to the introduce the hybrid equations of motion.] 
Thus, to get even a qualitatively accurate evolution of the binary
near coalescence, it may be necessary to use a full general relativistic
hydrodynamic treatment of the coalescence problem, and begin
the evolution when the stars are still well separated.

As we have shown, the dynamical instability in the equation
of motion will, in effect, cut off the chirping waveform.
The frequency of the cut-off is somewhat dependent on the
neutron star equation of state. Only in this late stage of the
evolution does the equation of state leave a tell-tale sign in the
emitted waveform. 
However, devising a strategy to dig this information
from the detector output requires further consideration.
Most detection/measurement strategies for coalescing binaries
involve integrating template waveforms against long stretches 
(8000 orbits!) of raw output data,
the idea being that one can detect/measure 
a relatively low amplitude signal by integrating for a long time.
Looking for the signature of this very late stage
plunge is precisely the opposite:
we are looking at the waveform just before coalescence
when the amplitude is fairly strong,
but the plunge is of fairly short in duration.
So answering questions about the plunge (such as, at what orbital
frequency did it begin?) requires measuring a relatively large
amplitude, but short-duration, effect.
Clearly, analysis of such events
will require a different detection strategy\cite{Kennefick96}.

\acknowledgments
We thank Kip Thorne for useful discussions. This work has been
supported by NSF Grants AST-9417371, PHY-9424337 and 
NASA Grant NAGW-2756 to Caltech. DL also
acknowledges support of the Richard C. Tolman Fellowship at Caltech.



\begin{figure}
\caption
{The critical orbital frequency (at the inner-most stable orbit)
as a function of the ratio of the neutron star radius $R_o$ and
mass $M$. The solid curves show the results including both 
relativistic and
tidal effects (the lower curve is for $\Gamma=3$ while the upper one 
is for $\Gamma=2$), the dashed curves are the Newtonian limit
given by Eqs.~(8)-(9). The vertical line corresponds to
$R_o/M=9/4$, the minimum value for any physical neutron star. 
The insert is a close-up for the nominal range of $R_o/M=4-8$
as given by all the available nuclear EOS's. 
Two curves within the insert should bracket all the physical 
values of $f_{\rm isco}$. 
}
\end{figure}

\begin{figure}
\caption
{The radial infall coordinate velocity during binary coalescence, with 
$M=M'=1.4M_\odot$, $R_o/M=5$, $\Gamma=3$, all calculated using 
$2.5PN$ radiation reaction.    
The solid line is the result including relativistic and tidal effects,
the short-dashed line includes only tidal effects, the long-dashed
line includes only relativistic effects.
The dotted line is the point mass ``Newtonian'' result. 
}
\end{figure}

\begin{figure}
\caption
{The number of orbits the binary spends per logarithmic
frequency. The labels are the same as in Fig.~2. 
}
\end{figure}

\begin{figure}
\caption
{The quadrupolar gravitational energy emitted near a given frequency.
The labels are the same as in Fig.~2.
}
\end{figure}

\begin{table}
\caption{Physical quantities at the inner-most stable orbit 
(dynamical stability limit) of neutron star binary,
with $M=M'$, $\Gamma=3$, and zero spin at large orbital radii.
Here $R_o$ is the neutron star radius, $a_1,a_2,a_3$ are the
axes of the ellipsoidal neutron star ($a_1$ along the binary axis,
$a_3$ perpendicular to the orbital plane), and
$M_{1.4}=M/(1.4M_\odot)$.}
The case in the top row is the purely Newtonian calculation using
LS equations of motion.
The last row is the point-mass calculation using the
hybrid equations of motion of KWW.
\begin{tabular}{c c c c c c}
$R_o/M$ & $r/R_o$ & $r/M_t$ & $a_2/a_1$ & $a_3/a_1$ &
$M_{1.4}f_{orb}$(Hz)\\
\hline
$---$ & $2.76$ & $6.90(R_o/5M)$ & $.772$ & $.805$ & 
$657(5M/R_o)^{3/2}$\\
$8$ & $2.87$ & $11.5$ & $.830$ & $.850$ & $279$\\
$6$ & $2.97$ & $8.91$ & $.857$ & $.871$ & $399$\\
$5$ & $3.10$ & $7.74$ & $.880$ & $.891$ & $488$\\
$4$ & $3.39$ & $6.78$ & $.915$ & $.921$ & $590$\\
$0$ & ---  & $6.03$ &  --- & --- & $697$\\
\end{tabular}
\end{table}


\begin{references}
\bibitem{notation}
We adopt units in which $G=c=1$. We use the following notation
to describe our binary system: The two neutron stars have (comparable)
masses $M$ and $M'$, the total mass of the system is denoted $M_t$; 
Small $r$ will denote orbital (center to center)
coordinate separation; capital $R_o$ will denote the neutron star 
radius at large $r$.
\end{references}
\end{document}